\documentclass[preprint,showpacs,floatfix]{revtex4}
\usepackage[dvips]{graphicx}
\begin{document}
\title{Why does the recently proposed simple empirical formula\\
for the lowest excitation energies work so well?}
\author{Guanghao \surname{Jin}}
\author{Dongwoo \surname{Cha}}
\author{Jin-Hee \surname{Yoon}}
\email{jinyoon@inha.ac.kr}
\thanks{Fax: +82-32-866-2452}
\affiliation{Department of Physics, Inha University, Incheon
402-751, Korea}
\date{January 28, 2008}

\begin{abstract}
It has recently been shown that a simple empirical formula, in terms of the mass number and the valence nucleon numbers, is able to describe the main trends of the lowest excitation energies of the natural parity even multipole states up to $10^+$ in even-even nuclei throughout the entire periodic table. In an effort to understand why such a simple formula is so capable, we investigate the possibility of associating each term of the empirical formula with the specific part of the measured excitation energy graph.
\end{abstract}


\maketitle

In a series of recent publications \cite{Ha1,Ha2,Yoon,Kim}, we have shown that a simple empirical formula was able to explain the main trends of the lowest excitation energies $E_x$ of the natural parity even multipole states in even-even nuclei throughout the entire periodic table. This formula, which depends only on the mass number $A$, the valence proton number $N_p$, and the valence neutron number $N_n$, is written as:
\begin{equation} \label{E}
E_x = \alpha A^{-\gamma} + \beta_p  e^{- \lambda_p N_p} + \beta_n
e^{- \lambda_n N_n}
\end{equation}
where the six model parameters, $\alpha$, $\gamma$, $\beta_p$, $\lambda_p$, $\beta_n$, and $\lambda_n$ are determined from the data for each multipole. The particular structure of Eq.\,(\ref{E}) was first envisaged by inspecting the plots of the measured lowest excitation energies, which were connected along the isotopic and isotonic chains \cite{Kim}. While this equation has a very simple form that can explain the main shape of the lowest excitation energy graph for all the natural parity even multipole states up to $10^+$ in even-even nuclei, any possible reasons why it is so capable have not been discussed until now. The intention of this study is therefore to investigate the possibility of associating each term of Eq.\,(\ref{E}) with some specific part of the measured excitation energy graph.

\begin{figure}[b]
\centering
\includegraphics[width=11.0cm,angle=0]{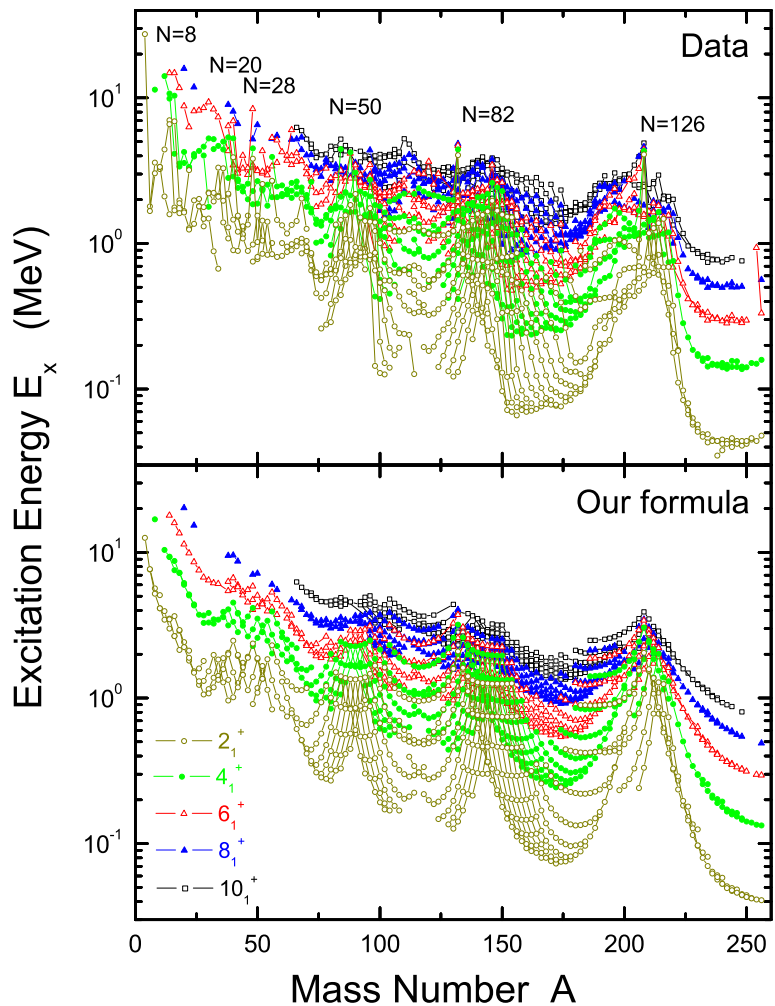}
\caption{Excitation energies of the first $2^+$ (empty circle), $4^+$ (solid circle), $6^+$ (empty triangle), $8^+$ (solid triangle), and $10^+$ (empty square) states in even-even nuclei. The points are connected by solid lines along the isotopic chains. The upper panel shows the measured excitation energies \cite{Raman,Firestone}, while the lower panel shows those calculated by Eq.\,(\ref{E}) with the parameter set given in Tab.\,\ref{tab-1}.}
\label{fig-1}
\end{figure}

In the upper panel of Fig.\,\ref{fig-1}, the measured excitation energies are shown of the first $2^+$ (empty circle), $4^+$ (solid circle), $6^+$ (empty triangle), $8^+$ (solid triangle), and $10^+$ (empty square) states in even-even nuclei as a function of the mass number, $A$. The measured excitation energies for the first $2^+$ states are cited from the compilation in Raman {\it et al}. \cite{Raman} and those for the other multipole states are extracted from the Tables of Isotopes, 8th edition by Firestone {\it et al}. \cite{Firestone}. The plotted points are connected by solid lines along the isotopic chains. The lower panel of Fig.\,\ref{fig-1} shows the energies of the first excited states calculated by Eq.\,(\ref{E}), with the parameter set given in Tab.\,\ref{tab-1}. The parameter values in Tab.\,\ref{tab-1} were calculated in Ref. \cite{Kim}, while the graph in Fig.\,\ref{fig-1} is actually the same as the graphs found in Figs.\,1-5 of Ref.\,\cite{Kim}. It is evident that the shape of the graph drawn from Eq.\,(\ref{E}) in the lower panel is quite similar to the shape of the graph that was determined by the data in the upper panel. Considering the fact that Fig.\,\ref{fig-1} contains all the lowest excitation energies of the natural parity even multipole states up to $10^+$ in all the even-even nuclei throughout the entire periodic table, the prediction power of Eq.\,(\ref{E}) should not be underestimated.

\begin{table}[b]
\begin{center}
\caption{The values adopted for the six parameters in Eq.\,(\ref{E}) for the excitation energy of the first natural parity even multipole states including $2_1^+$, $4_1^+$, $6_1^+$, $8_1^+$, and $10_1^+$ states. These parameter values have been quoted from Ref. \cite{Kim}. The last column is the total number, $N_{\rm tot}$ of the data points for the corresponding multipole state.}
\begin{tabular}{cccccccc}
\hline\hline
$J_1^\pi$~~~&~~~~~$\alpha$~~~~~&~~~$\gamma$~~~&~~~$\beta_p$~~~&
~~~$\beta_n$~~~&~~~$\lambda_p$~~~&~~~$\lambda_n$~~~&~~$N_{\rm tot}$~~\\
&(MeV)&&(MeV)&(MeV)&&&\\
\hline
$2_1^+$&68.37&1.34&0.83&1.17&0.42&0.28&557\\
$4_1^+$&268.04&1.38&1.21&1.68&0.33&0.23&430\\
$6_1^+$&598.17&1.38&1.40&1.64&0.31&0.18&375\\
$8_1^+$&1438.59&1.45&1.34&1.50&0.26&0.15&309\\
$10_1^+$&2316.85&1.47&1.36&1.65&0.21&0.14&265\\
\hline \hline
\end{tabular}
\label{tab-1}
\end{center}
\end{table}

\begin{figure}[b]
\centering
\includegraphics[width=11.0cm,angle=0]{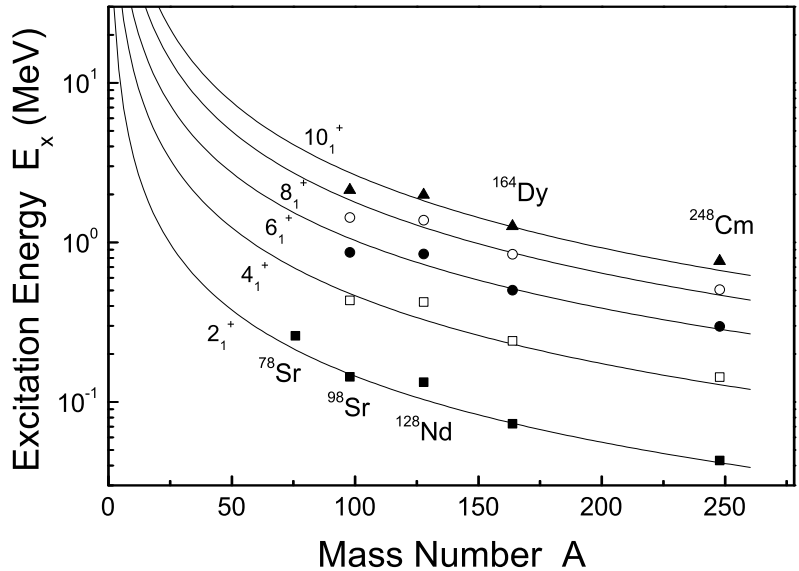}
\caption{Measured excitation energies of the first $2_1^+$ (solid square), $4_1^+$ (empty square), $6_1^+$ (solid circle), $8_1^+$ (empty circle), and $10_1^+$ (solid triangle) states of five nuclei that are near the doubly mid-shell region with large valence nucleon numbers. The solid lines are drawn according to Eq.\,(\ref{Emid}).}
\label{fig-2}
\end{figure}

\begin{figure}[b]
\centering
\includegraphics[width=11.0cm,angle=0]{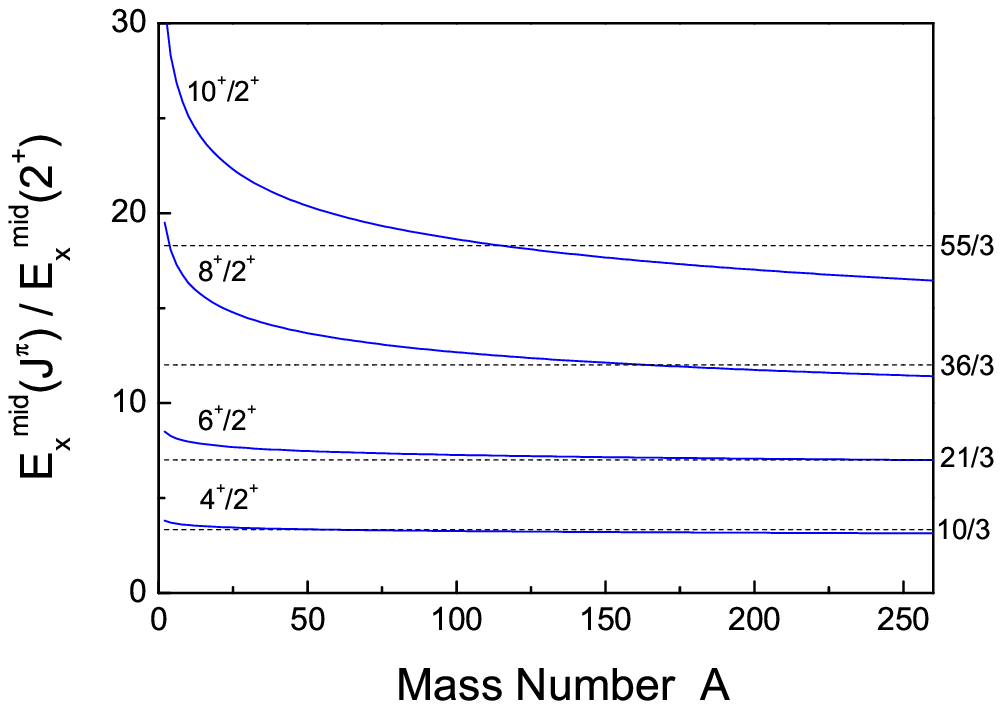}%
\caption{Excitation energy ratios $E_x^{\rm mid} (J^+) / E_x^{\rm mid} (2^+)$ (solid lines) calculated by the empirical formula, Eq.\,(\ref{Emid}). The horizontal dashed lines denote the same energy ratios determined by the rotational band energies, Eq.\,(\ref{Erot}).} \label{fig-3}
\end{figure}

The source of the beauty demonstrated by Eq.\,(\ref{E}) in its ability to reproduce such a complex excitation energy graph, as shown in the upper panel of Fig.\,\ref{fig-1}, can be traced back to the valence nucleon numbers $N_p$ and $N_n$ adopted in this equation. The valence proton (neutron) number $N_p$ ($N_n$) of a nucleus, with the atomic (neutron) number $Z$ ($N$), is defined by
\begin{equation} \label{v}
N_p \ (N_n) = \left\{ \begin{array}{ll}
   Z \ (N)-N_{c-1} &\quad \mbox{if $N_{c-1} < Z \ (N) \leq  M_c$}\\
   N_c - Z \ (N)  &\quad \mbox{if $M_c < Z \ (N) \leq N_c$}\end{array} \right.
\end{equation}
where, $N_c$ is the magic number for the $c$-th major shell, which is given by $N_0 = 0$, $N_1= 8$, $N_2= 20$, $N_3= 28$, $N_4= 50$, $N_5= 82$, and $N_6= 126$, and so on. Also, $M_c$ is the average of the two adjacent magic numbers, $(N_{c-1}+N_c)/2$, which corresponds to the number of nucleons contained in the mid-shell nucleus of the $c$-th major shell. The valence nucleon numbers, $N_p$ and $N_n$, are maximum for the mid-shell nuclei and zero for the magic shell nuclei. Eq.\,(\ref{v}) conveys that the valence proton (neutron) number $N_p$ ($N_n$) is equal to the number of proton (neutron) particles above the highest filled major shell, or is equal to the number of proton (neutron) holes, if the Fermi level is beyond the mid-shell within the highest proton (neutron) major shell. The valence nucleon numbers $N_p$ and $N_n$, for example, of a Mercury isotope $_{\ 80}^{208}{\rm Hg}_{128}$ are $N_p = 2$ and $N_n=2$ because there are two proton holes and two neutron particles in $_{\ 80}^{208}{\rm Hg}_{128}$.

The right hand side of Eq.\,(\ref{E}) is composed of three terms; the first term, which depends only on the mass number $A$, the second term, which depends only on the valence proton number $N_p$, and the third term, which depends only on the valence neutron number $N_n$. And it is extremely satisfying to realize that meaningful quantities, which are related exclusively to each term of Eq.\,(\ref{E}), can be extracted from the data shown in Fig.\,\ref{fig-1} by the method described below.

Firstly, by taking the lowest excitation energy $E_x^{\rm mid}$ found in the nucleus near the proton mid-shell and near the neutron mid-shell, the excitation energy can then be expressed by only the first term of Eq.\,(\ref{E}) as:
\begin{equation} \label{Emid}
E_x^{\rm mid} \approx \alpha A^{-\gamma}.
\end{equation}
This is because the remaining two terms of Eq.\,(\ref{E}) become practically zero, since the values of $N_p$ and $N_n$ are quite large and amount to several tens for nuclei which belong to higher major shells. Fig.\,\ref{fig-2} shows the measured excitation energies $E_x (2_1^+)$ (solid square), $E_x (4_1^+)$ (empty square), $E_x (6_1^+)$ (solid circle), $E_x (8_1^+)$ (empty circle), and $E_x (10_1^+)$ (solid triangle) of the lowest excited states found in five nuclei. These nuclei are near the doubly mid-shell region with large valence nucleon numbers such as $_{38}^{78}{\rm Sr}_{40}$, $_{38}^{98}{\rm Sr}_{60}$, $_{\ 60}^{128}{\rm Nd}_{68}$, $_{\ 66}^{164}{\rm Dy}_{98}$, and $_{\ 96}^{248}{\rm Cm}_{152}$. Fig.\,\ref{fig-2} also shows solid lines drawn according to Eq.\,(\ref{Emid}) that represent  five natural parity even multipoles, $2^+ \sim 10^+$. As expected, it can be seen in Fig.\,\ref{fig-2} that the measured excitation energies lie quite close to the lines predicted by Eq.\,(\ref{Emid}). Thus the two parameters, $\alpha$ and $\gamma$, from Eq.\,(\ref{E}) can be determined solely by the excitation energies found in nuclei near the doubly mid-shell region.

In addition, it is well known that the low-energy spectra of even-even nuclei, especially near the doubly mid-shell region, exhibit a strikingly simple pattern \cite{Bohr}. Such a pattern is known as the rotational band and consists of a sequence of states with $J^{\pi}=2^+$, $4^+$,$\cdots$, and their excitation energies $E_x^{\rm mid}(J^+)$ follow the relation
\begin{equation} \label{Erot}
E_x^{\rm mid} (J^+) = {{J(J+1) \hbar^2} \over 2I}
\end{equation}
where $I$ is the effective moment of inertia of the nucleus. The pattern can be clearly observed from the data in the upper panel of Fig.\,\ref{fig-1}, while the recently proposed empirical formula satisfactorily reproduces this pattern as can be seen from the lower panel of the same figure.

One of the characteristic features of the rotational band is the constancy of the energy ratios $E_x^{\rm mid} (J^+) / E_x^{\rm mid} (2^+)$ as determined by Eq.\,(\ref{Erot}), namely $E(J)/E(2)=10/3$, $21/3$, $36/3$, and $55/3$ for $J=4$, 6, 8, and 10, respectively \cite{Ring}. In Fig.\,\ref{fig-3}, the solid lines represent the energy ratios $E_x^{\rm mid} (J^+) / E_x^{\rm mid} (2^+)$ calculated by the empirical formula, Eq.\,(\ref{Emid}), while the horizontal dashed lines depict the same energy ratios given by the rotational band energies, Eq.\,(\ref{Erot}). From this figure, it can be seen that the first term $\alpha A^{-\gamma}$ of the recently proposed empirical formula complies with the rule of the rotational band for a wide range of the mass number $A$ and, therefore, we can say that the two parameters $\alpha$ and $\gamma$ in the recently proposed  empirical formula carry the information about the effective moment of inertia for deformed nuclei near the doubly mid-shell region.

The next consideration is the difference, $\Delta E_x^{\pi}$, of the excitation energies between the two adjacent nuclei in which neutron numbers differ by two along the isotopic chain. It then reduces to
\begin{equation} \label{E_pi}
\begin{array}{ll}
\Delta E_x^{\pi} = E_x (Z, N+2) - E_x (Z, N) \\
\quad \quad \ \, \approx \beta_n \left[ e^{-\lambda_n (N_n \pm 2)} - e^{-\lambda_n N_n} \right] \label{E_pi} \\
\quad \quad \ \, \approx \mp 2 \beta_n \lambda_n e^{-\lambda_n N_n}
\end{array}
\end{equation}
because $(A+2)^{-\gamma} \approx A^{-\gamma}$ for large $A$ and the second term of Eq.\,(\ref{E}) is constant for nuclei along the isotopic chain. The upper (lower) sign applies when the nucleus is located below (above) the neutron mid-shell. We can relate the parameter product $\beta_n\lambda_n$ directly to a measured quantity if we evaluate the difference $\Delta E_x^{\pi}$ for magic nuclei where $N_n=0$. Eq.\,(\ref{E_pi}) is then further simplified to
\begin{equation} \label{E_N}
\big| \Delta E_x^\pi (N_n=0) \big| \approx  2 \beta_n\lambda_n.
\end{equation}
Similarly, the excitation energy difference $\Delta E_x^{\nu}$ between the two adjacent nuclei in which atomic numbers differ by two along the isotonic chain can be simplified to
\begin{equation} \label{E_P}
\big| \Delta E_x^\nu (N_p=0) \big| \approx  2 \beta_p\lambda_p
\end{equation}
when evaluated for $N_p=0$.

\begin{table}[t]
\begin{center}
\caption{Average values of the excitation energy differences, $\big| \Delta E_x^\pi \big|$ and $\big| \Delta E_x^\nu \big|$ and the parameter products $2\beta_n\lambda_n$ and $2\beta_p\lambda_p$.}
\begin{tabular}{ccccc}
\hline\hline
~~$J_1^\pi$~~~&~~~$<\big| \Delta E_x^\pi \big|>$~~~&~~~~$2\beta _n \lambda _n$~~&~~~$<\big| \Delta E_x^\nu \big|>$~~&~~~~$2\beta _p \lambda _p$~~\\
&(MeV)&(MeV)&(MeV)&(MeV)\\
\hline
$2_1^+$&0.99$\pm$0.02&0.66&0.70$\pm$0.04&0.70\\
$4_1^+$&0.97$\pm$0.07&0.77&0.59$\pm$0.08&0.80\\
$6_1^+$&0.63$\pm$0.05&0.59&0.99$\pm$0.16&0.87\\
$8_1^+$&0.50$\pm$0.07&0.45&0.49$\pm$0.14&0.70\\
$10_1^+$&0.45$\pm$0.11&0.46&0.31$\pm$0.07&0.57\\
\hline \hline
\end{tabular}
\label{tab-2}
\end{center}
\end{table}

The average values of the left hand sides of Eq.\,(\ref{E_N}) and (\ref{E_P}) have been taken from the data. The results for the even multipole natural parity states up to $10^+$ are listed in Tab.\,\ref{tab-2} in the second (for $\big| \Delta E_x^\pi \big|$) and in the fourth (for $\big| \Delta E_x^\nu \big|$) columns. These averages can be compared with the parameter products listed in the third ($2\beta_n\lambda_n$) and in the last ($2\beta_p\lambda_p$) columns in Tab.\,\ref{tab-2}, where the parameter values are quoted from Tab.\,\ref{tab-1}. From Tab.\,\ref{tab-2} it can be seen that the $\big|\Delta E_x \big|$ averages taken from the data are reasonably represented by the product of parameters $\beta$ and $\lambda$, which are determined by the recently proposed empirical formula. Thus the two parameters $\beta_n$ and $\lambda_n$ in our empirical formula can be determined solely by the excitation energy difference between the two adjacent nuclei along the isotopic chain and the two parameters $\beta_p$ and $\lambda_p$ can be similarly determined along the isotonic chain.

As a first step in the effort to understand why the empirical formula, Eq.\,(\ref{E}), with such a simple form can describe the lowest excitation energy graph of natural parity even multipole states in all the even-even nuclei throughout the entire periodic table, we have shown that each term of Eq.\,(\ref{E}) can be associated with some specific part of the measured excitation energy graph. The excitation energies $E_x$, shown in Fig.\,\ref{fig-1}, can be divided into the following two parts: the first part is where $E_x$'s are minimum and the second part is where $E_x$'s are maximum. The excitation energies $E_x$ are minimum in the mid-shell nuclei where the valence nucleon numbers $N_p$ and $N_n$ are large. Therefore, $E_x$ can be represented by only the first term $\alpha A^{-\gamma}$ which is dependent on the mass number $A$ since values of the remaining exponential terms are practically zero due to the large negative exponent. Furthermore, the excitation energies of even-even nuclei near the doubly mid-shell region constitute the well-known rotational band in which energies are expressed simply by Eq.\,(\ref{Erot}). Therefore, the two parameters $\alpha$ and $\gamma$ in Eq.\,(\ref{E}) can be fixed solely by the excitation energies found in the nuclei near the doubly mid-shell region, and can carry the information about the effective moment of inertia for such nuclei. In the case where excitation energies $E_x$ are maximum, the difference is taken of the $E_x$'s between the two adjacent nuclei in which neutron numbers differ by two along the isotopic chain. We have then shown that the difference is only just equal to the parameter product, $2 \beta_n \lambda_n$. Furthermore, when the difference is taken of the $E_x$'s along the isotonic chain, the difference is equal to $2\beta_p \lambda_p$. The six parameters in the empirical formula can therefore be fixed exclusively by a different part of the excitation energy graph. It is therefore concluded that this is partly the reason why the recently proposed simple empirical formula works so effectively.

\begin{acknowledgments}
This work was supported by an Inha University research grant.
\end{acknowledgments}


\begin{references}
\bibitem{Ha1} E. Ha and D. Cha, J. Korean Phys. Soc. {\bf 50}, 1172 (2007).
\bibitem{Ha2} E. Ha and D. Cha, Phys. Rev. {\bf C}75, 057304 (2007).
\bibitem{Yoon} J-H Yoon, E. Ha, and D. Cha, J. Phys. G: Nucl. Part. Phys. {\bf 34}, 2545 (2007).
\bibitem{Kim} D. Kim, E. Ha, and D. Cha, Nucl. Phys. {\bf A}799, 46 (2008).
\bibitem{Raman} S. Raman, C. W. Nestor, Jr., and P. Tikkanen, At.
Data Nucl. Data Tables {\bf 78}, 1 (2001).
\bibitem{Firestone} R. B. Firestone, V. S. Shirley, C. M. Baglin, S. Y. F. Chu, and J. J. Zipkin, {\it Table of Isotopes}, 8th Edition (Wiley, New York, 1999).
\bibitem{Bohr} A. Bohr and B. R. Mottelson, {\it Nuclear Structure}, Vol. II (W. A. Benjamin, London, 1975).
\bibitem{Ring} P. Ring and P. Schuck, {\it The nuclear many-body problem}, (Springer-Verlag, New York, 1980).



\end{references}
\end{document}